\documentclass[11pt]{article}
\usepackage{graphicx}
\begin{document}
\def\<{\!<\!} \def\>{\!>\!} \def\={\!=\!} 
\def\+{\!+\!} \def\-{\!-\!} 
\setcounter{page}{0}
\setcounter{footnote}{0}
\begin{titlepage}
\vspace*{5mm}
\begin{flushright}
\small\sc  November 2000 \\
\small\sc  hep-th/0011004 \\
\end{flushright}
\vspace{20mm}
\begin{center}
{\Large {\bf Thermodynamic Bethe ansatz \\ [2mm]
   for generalized extensive statistics} }

\vspace{18mm}
 {\large Andrei G. Bytsko   }

\vspace{5mm}
{\em  Steklov Mathematics Institute\\
Fontanka 27, St.Petersburg, 191011, Russia }
\end{center}
\vspace{2.5cm}

\begin{abstract}
\noindent
We investigate properties of the entropy density related 
to a generalized extensive statistics and derive the 
thermodynamic Bethe ansatz equation for a system 
of relativistic particles obeying such a statistics.
We investigate the conformal limit of such a system. 
We also derive a generalized Y-system.
The Gentile intermediate statistics and the statistics 
of $\gamma$-ons are considered in detail. In particular,
we observe that certain thermodynamic quantities for
the Gentile statistics majorize those for the 
Haldane-Wu statistics. 
Specifically, for the effective central charges related
to affine Toda models we obtain nontrivial inequalities
in terms of dilogarithms.

\vspace*{20mm}

\par\noindent
\end{abstract}
\vfill{ \hspace*{-9mm}
\begin{tabular}{l}
\rule{6 cm}{0.05 mm}\\
bytsko@pdmi.ras.ru  \\
Research supported in part by INTAS grant 99-01459 and by
 RFFI grant 99-01-00101
\end{tabular}}
\end{titlepage}
\newpage \baselineskip 14.5pt 

\section{Introduction}
Although all experimentally observed particles are either bosons or
fermions, the general principles of quantum mechanics do not prohibit
existence of particles obeying other types of statistics \cite{GM}.
The motivation to consider an exotic statistics is that it may 
provide an {\em effective} description of the dynamics of particles 
if their interaction is so strong that the available occupancy for 
a given state depends on the number of particles already present in 
the state \cite{Ha}. An exotic statistics can also arise if particles 
possess a hidden internal degree of freedom that is invisible in the
Hamiltonian but that can become dynamically relevant.
Finally, an exotic statistics can emerge in description of physical
models in one or two spatial dimensions. The reason is that in this
case the multi-particle wave function is not necessarily even or odd
and particles (anyons) may obey a fractional statistics \cite{any}.
Moreover, in low-dimensional theories the exchange statistics of
the fields present in a Hamiltonian is not directly related to the
exclusion statistics of the corresponding particles. For instance,
although particles in the one-dimensional real coupling affine Toda
field theories (ATFT) are bosons, one has to impose the Pauli 
principle on their momenta in the thermodynamic Bethe ansatz analysis 
in order to obtain correct values of corresponding central charges 
\cite{TBAKM}. Another example is the conformal field theory, where 
quasi-particle spectra can be constructed with use of an exotic 
exclusion statistics \cite{quasi}.

The first attempt to introduce an exotic statistics is attributed to
G.~Gentile \cite{Gent}, who proposed an intermediate statistics,
in which at most $G$ indistinguishable particles can occupy a given 
state. This intermediate statistics interpolates between fermions and 
bosons, that are recovered for $G \= 1$ and $G \= \infty$.
A decade after Gentile's work H.S.~Green introduced the parafermi 
statistics \cite{Gr} that also fixes maximal occupancy for a given 
state. Since then, various aspects of systems governed by the Gentile 
or parafermi statistics were discussed in the literature, among them 
possible applications to the elementary particles theory \cite{gen} 
and to the quark confinement problem in the QCD \cite{qu}.

Many authors discussed thermodynamic properties of an ideal gas
obeying the Gentile or parafermi statistics \cite{Gent,gas,app}. 
In the present paper we will study the thermodynamic limit of 
one-dimensional relativistic integrable field models governed by a 
Gentile-like statistics. More precisely, the main purpose of this 
manuscript is to implement systematically a generalized extensive 
exclusion statistics into the thermodynamic Bethe ansatz (TBA) 
analysis of systems where the interaction of particles is 
relativistic, short-range and characterized by a factorizable 
scattering matrix.

The TBA was originally developed in the seminal papers by Yang 
and Yang \cite{Yang} for treating a one-dimensional ideal gas. 
In later works \cite{Zam,Y} this technique was adopted to 
one-dimensional relativistic models. The boundary condition for 
a many-particle wave function leads to what is commonly referred 
to as the Bethe ansatz equation, that provides the quantization 
condition for possible momenta of the system. The TBA constitutes 
an interface between massive integrable models and conformal 
field theories. From the TBA one can extract information about 
the ultraviolet limit of the underlying massive model, 
in particular, find the corresponding (effective) central charge.

In the derivation of the TBA equation the underlying exclusion
statistics is usually taken to be either of bosonic or fermionic 
type \cite{TBAKM,Yang,Zam}. In the present paper the TBA equation 
will be derived and studied for significantly more general situation.
Namely, let $W(N,n)$ denote the total dimension of the Hilbert space
for a system of $n$ indistinguishable particles that can occupy $N$
different states in the Fock-space. Assume that there exists a 
function such that its $N$-th power is
a generating function for the dimensions $W(N,n)$, i.e.
\begin{equation}\label{fgen}
 \bigl( f(t) \bigr)^N = \sum_{k \geq 0} W(N,k) \, t^k \,.
\end{equation}
In the statistical mechanics, if the variable $t$ is understood as
fugacity, the sum on the right hand side is the grand partition
function of the system. The property of a grand partition function
to be an $N$-th power of a function independent of $N$ is called
{\em extensivity}. As we are going to demonstrate below, this 
property allows us to develop the TBA analysis for a rather 
general choice of $f(t)$. Let us stress that for the purpose
of deriving the TBA equation even asymptotical extensivity suffices,
i.e., equation (\ref{fgen}) should hold in the large $N$ limit.

Of course, if the explicit form of $f(t)$ is known, we can use it to 
simplify the general formulae. For instance, an instructive example
is the Gentile statistics of order $G$ which is defined by the 
following choice of $f(t)$:
\begin{equation}\label{Gf}
 f_{\scriptscriptstyle G}(t) = 1+t+t^2+\ldots + t^G \,.
\end{equation}
For $G \= 1$ and $G \= \infty$ this statistics describes
ordinary fermions and bosons, respectively. Although 
$f_{\scriptscriptstyle G}(t)$ is 
the simplest non-trivial choice of $f(t)$, it has all features 
of a generalized extensive statistics. Therefore, the Gentile 
statistics will be our main working example below.

The paper is organized as follows.
In Section 2 we describe possible types of a generalized extensive
statistics and study properties of the corresponding entropy 
densities in the thermodynamic limit. The Gentile statistics and 
the $\gamma$-ons statistics are discussed in this context.
In Section 3 we compare properties of the Gentile statistics
and the Haldane-Wu statistics.
In Section 4 we derive the thermodynamic Bethe ansatz equation 
and the finite-size scaling function for a relativistic 
multi-particle system in which statistical interaction is 
governed by a generalized extensive statistics.
In Section 5 we derive the Y-system related to such a
generalized statistics.
In Section 6 we study the ultraviolet limit of the generalized
TBA equation and obtain an expression for the corresponding
effective central charge.
In Section 7 we compute finite-size scaling functions and
central charges related to some affine Toda models for the
Gentile statistics and the $\gamma$-ons statistics and compare 
them with their counterparts for the Haldane-Wu statistics. 
Our conclusions are stated in Section 8.

\section{Types of extensive statistics and entropy density}

Consider a system of $n$ indistinguishable particles that can occupy
$N$ different states in the Fock-space. Assume that the particles
obey the Gentile statistics of order $G$, i.e., that each state can
be occupied by at most $G$ particles (with $G$ being a positive
integer number). There are several combinatorial ways to compute
the total dimension of the Hilbert space for such a system.
For instance, we first choose $m_1$ states which are occupied by
at least one particle, then among these $m_1$ states we choose
$m_2 \leq m_1$ states which are occupied by at least two particles,
etc. This way of counting yields the following expression for
the total dimension of the Hilbert space
\begin{equation}\label{Gcount}
 W_G(N,n) = \sum_{N \geq m_1  \geq \ldots \geq m_{G-1}
 \geq 0}   C_N^{m_1} \, C_{m_1}^{m_2} \ldots C_{m_{G-2}}^{m_{G-1}}
 \, C_{\, m_{G-1}}^{n-m_1- \ldots - m_{G-1}} \,,
\end{equation}
where $C_k^m=k!/(m!(k-m)!)$ are binomial coefficients
(and $C_k^m=0$ if $m \> k$ or if $m \< 0$). Now, let us compute 
$N$-th power of the polynomial $f_{\scriptscriptstyle G}(t)$ 
defined in (\ref{Gf}). Thanks to the obvious recursive relation 
$f_{\scriptscriptstyle G}(t)=1+t\, f_{\scriptscriptstyle G-1}(t)$, 
we can do this by $G$ consecutive applications of the binomial 
formula. Then it is easy to see that $N$-th power of 
$f_{\scriptscriptstyle G}(t)$ is a generating function for the 
dimensions $W_G(N,n)$, i.e.
\begin{equation}\label{gen}
 \bigl( f_{\scriptscriptstyle G}(t) \bigr)^N  = 
  \sum_{n=0}^{G N} W_G(N,n) \, t^n \,.
\end{equation}
Thus, the Gentile statistics is a particular case of
an extensive statistics.

Let us stress that the order $G$ of the Gentile statistics must be
a positive integer. Although we could try to extend it to other
values by formal replacement of (\ref{Gf}) with
\begin{equation}\label{Gff}
 f_{\scriptscriptstyle G}(t) = \frac{1-t^{G+1}}{1-t} \,,
\end{equation}
the resulting function $f_{\scriptscriptstyle G}(t)$ will not 
satisfy the important positivity property (see below).

Consider now a more general system of $n$ indistinguishable
particles that possesses, at least for large $N$, the extensivity
property (\ref{fgen}) with
\begin{equation}\label{f}
   f(t) =  \sum_{k=0}^{d} P_k t^k \,,
\end{equation}
where $d$ can be a finite positive number or infinity.
The Taylor coefficients $P_k$ are fractional dimensions of levels
in a single-state Hilbert subspace and can be regarded as
probabilities of occupation of the given state by $k$ particles
(see related discussions in \cite{NW,Pn,Po}). The total dimension
of the single-state subspace is $f(1)$. It is natural to require
{\em positivity} of the quantities $P_k$:
\begin{equation}\label{posit}
 P_0=1 \qquad {\rm and} \qquad P_k \geq 0 \quad
      {\rm for} \quad k \geq 1 \,.
\end{equation}
The first condition here implies that the vacuum is realized with
probability one independently of the size $N$ of the system.
Furthermore, we assume that the series (\ref{f}) converges in the 
complex plane for $0 \leq |z| < R$, where $R \leq \infty$. This 
implies that $f(t)$ belongs to one of the following types:
\begin{equation}\label{types}
 \begin{array}{rl}
 {\rm I.} & \quad d < \infty \,, \qquad R=\infty \,; \\
 {\rm II.} & \quad  d = \infty \,, \qquad R=\infty \,; \\
 {\rm IIIa.} &\quad  d = \infty \,, \qquad 1< R < \infty \,; \\
 {\rm IIIb.} & \quad d = \infty \,, \qquad 0< R \leq 1 \,. 
 \end{array}
\end{equation}
The types I, II, and III consist of finite degree polynomials, 
analytic functions with infinite Taylor series, and meromorphic 
functions, respectively. For example, a finite order Gentile 
statistics belongs to the type I, the Boltzmann statistics 
($f(t) \= \exp t$) belongs to the type II, and the bosonic 
statistics ($f(t) \= (1 \- t)^{-1}$) is of the type IIIb. 
The reason we divided the type III into two subtypes will become 
clear below.

Let us remark that, in addition to (\ref{posit}), one may
require that $P_1 \= 1$ with the motivation that the statistics
should not be deformed if only one particle is present in the 
system. Although all the concrete cases which we consider below
do fulfill this requirement, it does not seem to be crucial for
the general considerations.

The entropy of the system under consideration is $S=k \ln W(N,n)$,
where $k$ is the Boltzmann constant. We need to compute the 
entropy in the thermodynamic limit, i.e., when $N\rightarrow \infty$ 
and $n/N$ is fixed. The extensivity property (\ref{fgen}) allows us 
to express $W(N,n)$ as the following contour integral
\begin{equation}\label{cint}
 W(N,n) = \frac{1}{2\pi i} \, \oint_{|z|=\rho} \, dz \,
 \bigl( f(z) \bigr)^N \, z^{-n-1}  =
 \frac{1}{2\pi i} \, \oint_{|z|=\rho} \frac{dz}{z} \,
 \exp \{ N ( \ln f(z) - \mu \ln z ) \} \,,
\end{equation}
where $\rho < R$, and we denoted $\mu = n/N$. Now we can find an
asymptotic expression for the entropy in the thermodynamic limit by
application of the saddle point method to the integral (\ref{cint}).
The saddle point of the exponent in (\ref{cint}), denote it $x$, 
is found as the positive
root of the equation (prime stands for a derivative)
\begin{equation}\label{xeq}
  x \, f^{\prime}(t)|_{t=x} = \mu \, f(x) \,.
\end{equation}
Choosing $\rho=x$ in (\ref{cint}), so that the saddle point
belongs to the integration contour, we apply the standard
result (see, e.g., \cite{asym}) for the Laplace integral:
\begin{equation}\label{expr}
 W(N,\mu N) =
  \frac{ \exp \{ N (\ln f(x) - \mu \ln x) \} }
       { \sqrt{ 2\pi N h_\mu(x) } } \,
  \Bigl( \frac{1}{x} + O(1/N) \Bigr)
 \qquad {\rm as} \quad N\rightarrow \infty \,,
\end{equation}
where $h_\mu (t)=f^{\prime\prime}(t)/(\mu f(t))+(1-\mu)/t^2$.
For any $f(t)$ satisfying (\ref{posit}) and $x$ obeying 
(\ref{xeq}) we have $h_\mu (x) \> 0$ due to the Hadamard 
theorem \cite{asym,Tit}. {}From (\ref{expr}) we obtain the 
entropy density $s(\mu)$ for a given value of $\mu$
\begin{equation}\label{lim}
 \lim_{N \rightarrow \infty} \frac{1}{N} \ln W(N,\mu N)
 \equiv s(\mu) = \ln f(x) - \mu \, \ln x \,,
\end{equation}
where $x$ is positive and satisfies (\ref{xeq}). 

Two remarks are in order here. Strictly speaking, if $f(t)$ 
is degenerated, i.e., if there exists integer $p>1$ such that 
$f(t) = \sum_{k} P_{pk} t^{pk}$, then the above derivation 
modifies since $f(t)$ has $p$ saddle points. However, the 
corresponding physical interpretation is simply that particles 
can be added only in $p$-tuples. Therefore, it is natural to 
introduce cluster variables $t^\prime = p t$, $n^\prime = p n$,
etc. In terms of these new variables formulae 
(\ref{cint})-(\ref{expr}) remain valid.

More important remark is that for deriving (\ref{lim}) the 
property $P_k \geq 0$ is crucial. In general, without the 
positivity property (\ref{cint}) does not have a real and uniform 
in $\mu$ limit. Moreover, if we allow $f(t)$ to have negative
Taylor coefficients, then some $W(N,n)$ in the expansion of 
$(f(t))^N$ may also be negative. This leads to an immediate 
problem with the definition of the entropy as a logarithm of 
$W(N,n)$. However, if the first negative coefficient in the 
expansion of $(f(t))^N$ appears only at sufficiently large 
power $n_0$ (of order $O(N)$), then $f(t)$ violating the 
positivity condition can still have well defined entropy 
density for certain range of $\mu$. For instance, such is the 
case of the Haldane-Wu 
statistics with statistical interaction $g$ (see Section 3), 
where $n_0 \approx N/g$. However, in the present paper we
will restrict our consideration, for simplicity, only to 
partition function strictly obeying the positivity condition.

The entropy density is a key quantity for developing the TBA 
analysis. {}From the definition of $\mu$ it is clear that 
$0 \!\leq\! \mu \!\leq\! d$ for $f(t)$ of the type I and 
$0 \!\leq\! \mu \!\leq\! \infty$ for the other types.
Formally this follows {}from equation (\ref{xeq}) which shows that
$\mu$ varies {}from $\mu(x \= 0)$ to $\mu(x \= R)$. Moreover,
$\mu(x)$ is a strictly increasing function because
$\partial_x \mu = x \mu h_\mu (x) \> 0$ for $x \> 0$. 

Since $f(t)$ is regular at zero, it is easy to see that $\mu \ln x$
tends to zero for small $\mu$. Together with the condition $P_0=1$
this yields $s(0)=0$. In order to analyze further behaviour of
$s(\mu)$, we employ (\ref{xeq}) and derive from (\ref{lim}) that
\begin{equation}\label{der}
 \partial_\mu s(\mu) = -\ln x \,.
\end{equation}
Since $\partial_x \mu \> 0$, we conclude that 
$\partial^2_\mu s(\mu) \< 0$, that is $s(\mu)$ is a convex up 
function. Furthermore, (\ref{der}) implies that 
$\max s(\mu) = \ln f( \min \{1,R\} )$. Note, however, that
$s(\mu)$ is not guaranteed to be positive for the whole range of 
$\mu$. More precisely, $s(\mu)$ may have one root at the interval
$\mu(1) < \mu \leq \mu(R)$. Indeed, for $f(t)$ of the type I we
derive {}from (\ref{xeq}) and (\ref{lim}) that
\begin{equation}\label{sd}
 s(d)=\ln P_d  \,.
\end{equation}
That is, $s(\mu)$ is non-negative everywhere at
$0 \!\leq\! \mu \!\leq\! d$ only if the last Taylor coefficient
$P_d$ is greater or equal to one.
For the other types of $f(t)$ we have the following properties
\begin{eqnarray}
 & s(\infty) = \infty \quad 
   \hbox{for type IIIb} \,, & \label{sinf1} \\
 & s(\infty) = -\infty \quad \hbox{\rm for types II and IIIa}
 \,. & \label{sinf2}
\end{eqnarray}
Indeed, in the first case $x \!\leq\! R \!\leq\! 1 $, hence, 
according to (\ref{der}), $s(\mu)$ is strictly increasing. 
In the second case we have 
$s(\infty)=\ln f(1)-\int_{\mu(1)}^{\infty} {\rm d}\mu \,\ln x$.
The integral here apparently diverges since it exceeds
$\int_{\mu(R_0)}^\infty {\rm d}\mu \, \ln R_0$, where we can
choose any $R_0$ such that $1 \< R_0 \< R$.

Investigation of thermodynamic properties of models in which the
entropy density becomes negative is rather problematic. Therefore
eqs.~(\ref{sd})-(\ref{sinf2}) suggest that we should restrict the 
choice of $f(t)$ to the type I with $P_d \geq 1$ 
(the Gentile-like statistics) and to the type IIIb
(the bose-like statistics). However, let us notice that in the
TBA framework the variable $x$ depends on the rapidity $\theta$.
For some relativistic models with factorizable scattering the
typical picture \cite{TBAKM} is that $x(\theta)$ falls off as
$|\theta|$ grows (for instance, $x(\theta)=\exp\{-mr \cosh \theta\}$
for the trivial S-matrix). In this case $x(\theta)<x(0)$ and we 
can consider $f(t)$ of the types II and IIIa if $s(\mu)$
remains non-negative for $\mu \leq \mu(x(0))$.

In the context of this discussion it is instructive to consider the
so-called $\gamma$-ons \cite{al,app}. These are particles with 
statistics interpolating between fermions ($\gamma \= 1$) and 
bosons ($\gamma \= -1$) in the following simple way
\begin{equation}\label{mga}
  \mu(x) = \frac{x}{1+\gamma x} \,.
\end{equation}
Solving equation (\ref{xeq}) for this choice of $\mu$ it is easy
to find the corresponding single-state partition function:
\begin{equation}\label{fga}
  f_\gamma(t) = (1+\gamma t)^{1/\gamma} =
 1 + t +  \sum_{k\geq 2} [k \- 1]_\gamma \frac{t^k}{k!} \,,
\end{equation}
where $[k]_\gamma \= \prod_{m=1}^{k} (1 \- m\gamma)$.
For $\gamma \> 0$ the positivity condition (\ref{posit}) is 
fulfilled only if $\gamma \= 1/d$, with $d$ positive integer. 
In this case $f_\gamma(t)$ is of the type I, the corresponding 
maximal occupancy is $\mu \= d$. Since $\ln P_d \< 0$ for 
$d \> 1$, the entropy density becomes negative for certain range 
of $\mu$. The case of $\gamma \= 0$ describes the Boltzmann 
statistics. Here $f_\gamma(t) \= \exp{t}$ belongs to the 
type II and $s(\mu)$ is negative for $\mu \> e$.
Finally, for $\gamma \< 0$ the positivity condition 
(\ref{posit}) is always fulfilled and $f_\gamma(t)$ is of the 
type IIIa or IIIb depending on the value of $\gamma$.

Returning to the Gentile statistics, we can summarize the properties
of the corresponding entropy density $s_{\scriptscriptstyle G}(\mu)$
as follows (see Fig.\ 1 for illustration).
$s_{\scriptscriptstyle G}(\mu)$ is a convex up function defined
for $0 \!\leq\! \mu \!\leq\! G$ and vanishing at the end points
of this interval. It attains the maximum at $\mu=G/2$:
\begin{equation}\label{Gmax}
  s_{\scriptscriptstyle G} (G/2) =  \ln (G+1) \,.
\end{equation}
Furthermore, $s_{\scriptscriptstyle G}(\mu)$ possesses the 
following symmetry
\begin{equation}\label{sym}
  s_{\scriptscriptstyle G} (G-\mu) =
  s_{\scriptscriptstyle G} (\mu) \,.
\end{equation}
This is a consequence of the equality $W_G(N,GN \- n)= W_G(N,n)$
which, in turn, follows {}from the relation (\ref{gen}).

\section{Comparison with Haldane-Wu statistics}

It is interesting to compare thermodynamic properties of a system
of particles obeying the Gentile statistics with those of a system
obeying the so-called Haldane-Wu statistics. For the latter, the
total dimension of the Hilbert space is postulated to be \cite{Wu}
\begin{equation}\label{Wahr}
 \hat{W}_g(N,n) = 
   \frac{(N+ (1-g)n +g-1)!}{n! \, ( N - g n + g-1)!} \,.
\end{equation}
This expression interpolates between the bosonic ($g=0$) and the
fermionic ($g=1$) statistics. Introduction of such an interpolating
statistics was motivated by Haldane's generalization \cite{Ha} of
the Pauli exclusion principle to the form
\begin{equation}\label{Pauli}
 \Delta D/\Delta n=-g \,.
\end{equation}
Here $D$ is the number of available states (holes) before the
$n$-th particle has been added to the system. The parameter $g$
is called the statistical interaction.
Properties of an ideal gas obeying the Haldane-Wu statistics
were actively discussed in the literature \cite{NW,Wu,HW}. The
corresponding TBA equation for relativistic integrable models was
obtained in \cite{BF} and investigated in \cite{BF,F}.

A direct comparison of the quantities $W_G(N,n)$ and $\hat{W}_g(N,n)$
is problematic if the system contains a finite number of particles.
Indeed, formula (\ref{Gcount}) has a bulky form for generic $G$
(i.e., for $2 \!\leq\! G \< \infty$).  Moreover, it requires
additional conventions to make exact sense of expression (\ref{Wahr})
for generic $g$ (i.e., for $0 \< g \< 1$). Actually, there
exists no prescription for counting of states on the microscopical
level that would lead to (\ref{Wahr}) (see related discussions in
\cite{NW,Pn,Po}). To overcome these difficulties we can compare the
two statistics in the thermodynamic limit. More precisely, let us
compare the entropy density $s_{\scriptscriptstyle G}(\mu)$ with its
counterpart $\hat{s}_g(\mu)\equiv \lim_{N \rightarrow\infty}
   \frac{1}{N} \ln\hat{W}_g(N,\mu N)$
that is readily derived from (\ref{Wahr}) with the help of the 
Stirling formula
\begin{equation}\label{limhw}
 \hat{s}_g(\mu)  = (1+\mu(1-g))\ln (1+\mu(1-g)) -
 \mu\ln \mu - (1-g\mu) \ln (1-g\mu) \,.
\end{equation}

Let us notice that the Haldane-Wu statistics is extensive but the
corresponding function $f_g(t)$ does not satisfy the positivity 
property (\ref{posit}). Indeed, we can reconstruct $f_g(t)$ 
first as a function of $\mu$ by the following formula
\begin{equation}\label{le}
 f(\mu) = \exp \{ s(\mu) - \mu \, \partial_\mu s(\mu) \} \,,
\end{equation}
that follows {}from (\ref{lim}) and (\ref{der}). 
For $\hat{s}_g(\mu)$ given by (\ref{limhw}) this yields
\begin{equation}\label{fhw}
 f_g(\mu) = \frac{1+(1-g)\mu}{1-g\mu} \,,
\end{equation}
which is a well-defined function. But reexpressing it in terms of
$t$ with the help of (\ref{xeq}) we will always obtain a function
that breaks the positivity property (\ref{posit}). For instance,
for $g=1/2$ we find (which coincides with the result of \cite{NW})
\hbox{$f_g(t) = 1 \+ t^2/2 \+ t \sqrt{1 \+ t^2/4} =$} 
\hbox{ $ 1 \+ t \+ t^2/2 \+ t^3/8 \- t^5/128 \+ \ldots$}.
However, the first negative term in expansion of the $N$-th power 
of this series appears for $n= 2N\+ 3$. This implies that in the
large $N$ limit thermodynamic quantities are well-defined for
$\mu \leq 2$.

For our purposes we can regard (\ref{limhw}) as {\em a definition} 
of the Haldane-Wu statistics. Then we have a finite maximal 
occupancy $N/g$ for a given number of particles $N$ simply because 
$\hat{s}_g(\mu)$ is defined for $0 \!\leq\! \mu \!\leq\! 1/g$. 
Furthermore, $\hat{s}_g(\mu)$ is a 
convex up function vanishing at the end points of this interval.
Thus, we see that properties of the entropy density 
$\hat{s}_g(\mu)$ are similar to those of a type I extensive 
statistics and, specifically, to those of the Gentile statistics. 
Therefore, it is natural to ask how much the behaviour of 
$s_{\scriptscriptstyle G}(\mu)$ differs {}from $\hat{s}_g(\mu)$ 
if $g=1/G$ (when the corresponding maximal occupancies coincide).

First, notice that the function 
$\hat{s}_{\scriptscriptstyle 1/G}(\mu)$ cannot coincide with 
$s_{\scriptscriptstyle G}(\mu)$ identically for generic $G$ 
since $\hat{s}_g(\mu)$ does not possess a symmetry like 
(\ref{sym}) if $g\neq 0,1$. Next, we can compare the 
mid-point value
$\hat{s}_{\scriptscriptstyle 1/G}(G/2)=\frac{G+1}{2}\ln (G+1) -
 \frac{G}{2}\ln G$
with the corresponding value of $s_{\scriptscriptstyle G}(\mu)$
given by (\ref{Gmax}). Taking into account the inequalities
\begin{equation} \label{Gin}
\begin{array}{l}
 t \ln t > (t-1)\ln(t+1) \qquad {\rm for} \quad t>1 \,, \\[0.5mm]
 t \ln t < (t-1)\ln(t+1) \qquad {\rm for} \quad 0<t<1 \,,
\end{array}
\end{equation}
we establish that
$ s_{\scriptscriptstyle G}(G/2) >
  \hat{s}_{\scriptscriptstyle 1/G}(G/2) $ for $G \> 1$.
Actually, numerical computations for various values of 
$G \> 1$ show that $s_{\scriptscriptstyle G}(\mu)$ 
majorizes $\hat{s}_{\scriptscriptstyle 1/G}(\mu)$
{\em everywhere} at $0 \< \mu \< G$,
\begin{equation}\label{ineq2}
 s_{\scriptscriptstyle G}(\mu) >
 \hat{s}_{\scriptscriptstyle 1/G}(\mu) 
 \qquad {\rm for} \quad G > 1 \,.
\end{equation}
For illustration, the case of $G=2$ is presented in Fig.~1.

Thus, we see that for generic $G$ the difference between
$\hat{s}_{\scriptscriptstyle 1/G}(\mu)$ and
$s_{\scriptscriptstyle G}(\mu)$ is not small if $\mu$ is not close
to $\mu=0$ or $\mu=G$. It is instructive to compare these functions
also near the end points (in particular, this will provide 
an additional support to the assertion (\ref{ineq2})).

In a vicinity of $\mu=0$ equation (\ref{xeq}) is solved as
$x=\mu +(2\delta_{G,1} -1)\mu^2 + O(\mu^3)$, and we find that
$  s_{\scriptscriptstyle G}(\mu) =
 (1-\ln\mu) \mu +(1/2-\delta_{G,1}) \mu^2 + O(\mu^3)  $,
where $\delta_{m,n}$ stands for the Kronecker symbol. Comparing
this expansion with
$  \hat{s}_g(\mu) = (1-\ln\mu) \mu + (1/2-g)\mu^2 + O(\mu^3) $
which follows {}from (\ref{limhw}), we infer that for small $\mu$
functions $\hat{s}_g(\mu)$ and $s_{\scriptscriptstyle G}(\mu)$
take close values (up to the order $\mu^2$) for {\em any choice} of
$G$ and $g$. Actually, employing (\ref{xeq}), it is easy to derive
the following expansion for the entropy density of an arbitrary
extensive statistics with $f(t)$ satisfying (\ref{posit}):
\begin{equation}\label{exp}
 s(\mu) = (P_1 - \ln \mu) \, \mu + O(\mu^2) \,.
\end{equation}
Thus, for small $\mu$ in the thermodynamic limit, the Haldane-Wu
and Gentile statistics are close not only to each other but to any
extensive statistics for which two first Taylor coefficients
of $f(t)$ are $P_0 = P_1 =1$. For instance, for small $\mu$ 
the Haldane-Wu and Gentile statistics are close also to  the 
statistics of $\gamma$-ons (\ref{fga}).

Consider now $\mu=G-\epsilon$ for small positive $\epsilon$. Then, 
employing (\ref{exp}) and the symmetry (\ref{sym}), we derive:
$  s_{\scriptscriptstyle G}(G-\epsilon) =
 (1-\ln\epsilon) \,\epsilon + O(\epsilon^2) $.
On the other hand, setting $\mu=1/g-\epsilon$ in (\ref{limhw}),
we obtain
$ \hat{s}_g(1/g-\epsilon) =
 (1-2\ln g -\ln\epsilon) \,g\,\epsilon +O(\epsilon^2) $.
We see that in a vicinity of $\mu=G$ the difference between
$s_{\scriptscriptstyle G}(\mu)$ and
$\hat{s}_{\scriptscriptstyle 1/G}(\mu)$ is not even of first 
order in $\epsilon$ but of order $\epsilon\ln\epsilon$
(so that the corresponding plots are visibly different here,
see Fig.~1).

Summarizing, it appears plausible that the Gentile statistics
of order $G>1$ {\em majorizes} the Haldane-Wu statistics with
parameter $g=1/G$ and these statistics are {\em not close}
except for small values of $\mu \= n/N$.

\section{Thermodynamic Bethe ansatz equation}
Now we will consider a relativistic multi-particle system of $l$
different species of particles confined to a finite region of the 
size $L$. We denote by $n_a$ the number of particles and by $N_a$ 
the dimension of the Fock-space related to the species $a$.
In the thermodynamic limit both the dimensions of the Fock-space
and the size $L$ of the region were the system is confined approach
infinity but the ratios $n_a/L$ remain finite \cite{Yang}.
For the fraction of particles with rapidities between
$\theta \- \Delta \theta /2$ and $\theta \+ \Delta \theta /2$
it is convenient to introduce densities
\begin{equation}\label{dens}
\Delta N_a = \rho_a(\theta )\,\Delta \theta \,L  \,,\qquad
\Delta n_a = \rho_a^{r}(\theta )\,\Delta \theta \,L  \,.
\end{equation}
As usual, the rapidity $\theta $ parameterizes the two-momentum
$P=m\left( \cosh \theta ,\sinh \theta \right) $ and
the total energy of the system is given by
\begin{equation}\label{ener}
 E \left[ \rho^r \,\right] = L  \sum_{a=1}^l
 \int_{-\infty}^{\infty}  {\rm d}\theta \,
 \rho_a^r (\theta ) \, m_a \cosh \theta \,.
\end{equation}
According to (\ref{dens}) the variable $\mu$ in (\ref{lim}) is
now related to the particle densities
\begin{equation}\label{mu}
  \mu_a (\theta)  = \rho_a^r (\theta)/\rho_a (\theta) \,.
\end{equation}
Thus, we can express the entropy as the following functional
\begin{equation}\label{entro}
 S[ \rho ,\rho^r \,] =   kL \sum_{a=1}^l
    \int_{-\infty }^{\infty }  {\rm d}\theta \, 
      \rho_a(\theta)  s(\mu_a)
  = kL \sum_{a=1}^l \int_{-\infty }^{\infty } {\rm d}\theta \, 
 \bigl\{ \rho_a(\theta) \ln f_a(x_a(\theta)) - 
  \rho_a^r(\theta) \ln x_a(\theta) \bigr\} \,,
\end{equation}
where each $x_a(\theta)$ satisfies (\ref{xeq}) and hence it is 
a function of $\rho_a^r (\theta)/\rho_a (\theta)$.
The subscript of $f_a$ means that different species may have
different single-state partition functions.

In the formulation of the TBA equation for ordinary statistics
it is common to introduce the  so-called pseudo-energies
$\epsilon_a(\theta)$ such that bosonic (upper sign) and 
fermionic (lower sign) distributions are
$ \rho^r_a(\theta)/\rho_a(\theta) = 
 \left( \exp (\epsilon_a (\theta))\mp 1 \right)^{-1}$.
In view of (\ref{xeq}), it is natural to define the 
pseudo-energies for an arbitrary extensive statistics as follows
\begin{equation}\label{Id1}
  \epsilon_a (\theta ) = - \ln x_a (\theta) \,,
\end{equation}
so that $ \rho^r_a(\theta)/\rho_a(\theta) = 
 -\partial_{\epsilon_a(\theta)} 
 \ln f_a( e^{-\epsilon_a(\theta)})$. 
For instance, the distribution for the Gentile statistics acquires 
the following form in terms of the pseudo-energies (with a slightly
different meaning it was introduced by Gentile \cite{Gent} for
an ideal gas)
\begin{equation}\label{Id2}
 \frac{\rho^r_a (\theta) }{ \rho_a (\theta) }=
 \frac{ \sum_{k=1}^{G_a} k \exp\{ (G_a-k)\epsilon_a (\theta) \} }{
 \sum_{k=0}^{G_a} \exp\{ k \epsilon_a (\theta) \} } \,.
\end{equation}

According to the fundamental principles of thermodynamics the
equilibrium state of a system is found by minimizing the free
energy $F$. Hence, keeping the temperature constant we obtain
the equilibrium condition by minimizing
$F\left[ \rho ,\rho^r \,\right] =E\left[ \rho^r \,\right]
-TS\left[ \rho ,\rho^r \,\right] $ with respect to $\rho_a^r$.
The equilibrium condition reads
\begin{equation}\label{equ}
 \frac{\delta F}{\delta \rho^r_a } =
 \frac{\delta E}{\delta \rho^r_a}
 - T \frac{\delta S}{\delta \rho^r_a}-
 T \sum_{b=1}^{l}\frac{\delta S}{\delta \rho_b }
 \frac{ \delta \rho_b }{ \delta \rho^r_a }=0    \,.
\end{equation}

The admissible momenta in the system are restricted by the boundary 
conditions. To derive the corresponding quantization equations one
takes a particle in the multi-particle wave function on a trip 
through the whole system \cite{Yang,Zam}. The particle will scatter 
with all other particles, which yields the following set of 
equations determining admissible rapidities 
\begin{equation}\label{BA}
 \exp (i L m_i \sinh \theta_i )\prod_{j\neq i}^{N}S_{ij}
 (\theta_i - \theta_j )= \kappa_i \,, \qquad i=1,\ldots,N \,,
\end{equation}
where $N \= \sum\nolimits_a N_a$, and $S_{ij}(\theta)$ is a 
two-particle scattering matrix. The constant phases $\kappa_i$ 
on the r.h.s.~of (\ref{BA}) may depend on the particle's species;
for our purposes their exact values are irrelevant. Taking the 
logarithmic derivative of (\ref{BA}) and employing densities as 
in (\ref{dens}), we obtain 
\begin{equation}\label{Bou}
 m_a \cosh \theta +  2\pi  \sum_{b=1}^l \, 
 \bigl( \varphi_{ab} * \rho^r_b \bigr) (\theta)
 = 2\pi \, \rho_a (\theta ) \,.
\end{equation}
Here $(u*v) (\theta) \equiv 1/(2\pi )\int {\rm d}\theta^\prime
 u(\theta -\theta^\prime) v(\theta^\prime)$ stands for the 
convolution and we introduced as usual the notation
$\varphi_{ab} (\theta)= -i\partial_\theta ( \ln S_{ab}(\theta) )$.

Employing equations (\ref{xeq}), (\ref{lim}) and (\ref{mu}), 
we compute variations of the entropy (\ref{entro}):
\begin{equation}
  \delta S/\delta \rho^r_a = -\ln x_a (\theta) \,, \qquad
  \delta S/\delta \rho_a = \ln f_{a} (x_a(\theta)) \,.
\end{equation}
Substitution of these relations into the equilibrium condition
(\ref{equ}) together with (\ref{Bou}) yields the desired 
thermodynamic Bethe ansatz equations for a system in which 
statistical interaction between particles of $a$-th species is 
described by a single-state partition function $f_a$,
\begin{equation}\label{GTBA}
 \frac{1}{kT} m_a \cosh \theta = 
 - \ln x_a (\theta) + \sum_{b=1}^l \,
 \bigl( \varphi_{ab} * \ln f_b (x_b) \bigr) (\theta) \,.
\end{equation}

In general, this set of integral equations cannot be solved 
analytically. However, one can try to solve these TBA equations
numerically with the help of the iteration method (as we will
do for some examples in Section 7). For the ordinary statistics
and the Haldane-Wu case it is known that such an iterative
procedure converges provided that 
$\varphi_{ab} (\theta)$ falls off sufficiently fast. 
The most complete proof of this assertion is presented in 
\cite{F,CK} and is based on the fixed point theorem. Presumably 
this proof extends to the case of (\ref{GTBA}) (although some 
further restrictions on $f_a(t)$ may be required), but we will 
not discuss this here. Let us notice only that if (\ref{GTBA}) 
can be solved iteratively and $\varphi_{ab}(\theta)$ is 
symmetric in $\theta$ (which is always the case for a unitary 
S-matrix), then $x_a(\theta)$ is also symmetric in $\theta$. 

We now substitute the expressions for the total energy 
(\ref{ener}) and the entropy (\ref{entro}) together with 
equations (\ref{Bou}) and (\ref{GTBA}) back into the expression 
for the free energy, and obtain 
\begin{equation}\label{Free}
 F(T)=-\frac{LkT}{2\pi } \sum_{a=1}^l m_a  \int_{-\infty}^{\infty}
 {\rm d}\theta \, \cosh \theta \, \ln f_{a} (x_a (\theta)) \, .
\end{equation}
The relation between the free energy and the finite-size scaling
function is well-known to be $c(T)=-6F(T)/(\pi LT^{2})$ \cite{Cardy}.
As usual we introduce a variable $r=1/T$ and set now the Boltzmann
constant to be one. Then the finite-size scaling function
acquires the form
\begin{equation}\label{c(r)}
 c(r)= \sum_{a=1}^l \frac{6 m_a r}{\pi^2} \int_0^{\infty}
 {\rm d}\theta \, \cosh \theta \, \ln f_{a} (x_a (\theta)) \,.
\end{equation}
Here we assumed that $x_a(\theta) \= x_a(-\theta)$. Once more 
for $f(t)=1+t$ and $f(t)=1/(1-t)$ we recover the well-known 
expressions for the fermionic and bosonic scaling functions.

It is instructive to compare the above TBA equation (\ref{GTBA}) 
and the finite-size scaling function (\ref{c(r)})
with those for the Haldane-Wu statistics \cite{BF}:
\begin{equation}\label{hwTBA}
 \frac{1}{kT} m_a \cosh \theta = \ln(1+y_a (\theta)) + 
 \sum_{b=1}^l \, \bigl( \Phi_{ab} * \ln (1+y^{-1}_b) 
 \bigr) (\theta) \,,
\end{equation}
\begin{equation}\label{hwcr}
 c_g(r) = \sum_{a=1}^l \frac{6 m_a r}{\pi^2} \int_0^{\infty}
 {\rm d}\theta \, \cosh \theta \, \ln (1+y^{-1}_a (\theta)) \,.
\end{equation}
Here $\Phi_{ab} (\theta) \= \varphi_{ab}(\theta) - 2\pi 
 g_{ab} \delta(\theta)$, and $g$ is a matrix which appears on
the r.h.s.~of (\ref{Pauli}) if we consider a system of several
species. As was discussed in \cite{BF}, the TBA equation
(\ref{hwTBA}) leads to an equivalence principle: two systems 
having the same mass spectra and identical quantities 
$\Phi_{ab}(\theta)$ are thermodynamically equivalent (as seen 
{}from (\ref{hwcr}), their finite-size scaling functions 
coincide). Our TBA equation (\ref{GTBA}) does not possess 
such a feature. More precisely, for (\ref{GTBA}) there exists 
no way to compensate a difference in statistics by a change of 
$\varphi_{ab}(\theta)$ independent of the explicit form of 
the S-matrix.

\section{Y-systems}

For some class of models it is possible to carry out certain 
manipulations on the TBA equations such that the original integral 
TBA equations acquire the form of a set of functional equations 
in new variables $Y_a$ \cite{Y}. These functional equations 
(commonly referred to as \hbox{Y-systems}) have the further 
virtue that unlike the original TBA equations they {\em do not} 
involve the mass spectrum. In the $Y$-variables certain periodicities 
in the rapidities are exhibited more clearly. These periodicities 
may then be utilized in order to express the quantity $Y_a(\theta)$ 
as a Fourier series, which in turn is useful to find solution of 
the TBA equations and expand the scaling function as a power series 
in the scaling parameter $r$. We will now demonstrate that similar 
equations may be derived for a multi-particle system in which the 
dynamical scattering is governed by the scattering matrix related 
to the ADE-affine Toda field theories and the statistical 
interaction is of a general type considered above.

Consider the minimal part (i.e., independent of the coupling 
constant) of the scattering matrix of the ADE-affine Toda field 
theories. As was shown in \cite{Rav}, upon appropriate choice 
of CDD-ambiguities, these S-matrices satisfy the identity
\begin{equation}
 \varphi_{ab} \Bigl( \theta +\frac{i\pi }{h} \Bigr) +
 \varphi_{ab} \Bigl( \theta -\frac{i\pi }{h} \Bigr)
 =\sum_{c=1}^{l} I_{ac} \varphi_{cb} ( \theta ) -
 2\pi I_{ab} \delta ( \theta ) \, ,
\label{RAD}
\end{equation}
where $h$ denotes the Coxeter number, $l$ the rank and $I$ the
incidence matrix of the corresponding Lie algebra. It is then 
straightforward to derive the ``Y-system''
\begin{equation}\label{YS}
 Y_a \Bigl( \theta +\frac{i\pi }{h} \Bigr) \,
 Y_a \Bigl( \theta -\frac{i\pi }{h} \Bigr)
 =\prod_{b=1}^{l} \Bigl( z_b(Y_b (\theta)) \Bigr)^{I_{ab}} \,,
\end{equation}
where $Y_a(\theta)=x_a^{-1}(\theta)$ and
\begin{equation}\label{Y}
 z_a(t) = t \, f_a(t^{-1}) \,.
\end{equation}
Equations (\ref{YS}) follow upon first adding (\ref{GTBA}) at
$\theta + \frac{i\pi }{h}$ and $\theta - \frac{i\pi }{h}$ and 
subtracting $I$ times (\ref{GTBA}) at $\theta $ from the sum. 
Thereafter we employ the fact \cite{Mass} that the masses of 
an ADE affine Toda field theory are proportional to the 
Perron-Frobenius vector of the corresponding Cartan matrix, 
i.e., $\sum_b C_{ab} m_{b} =4\sin^{2} (\pi/(2h)) m_a$.
Then, with the help of (\ref{RAD}), equations
(\ref{YS}) follow.

In comparison with (\ref{GTBA}) equations (\ref{YS}) have already
the virtue that they are simple functional equations and do not 
involve the mass spectrum. For the Gentile statistics we have 
$z(t)= t + 1 + t^{-1} + \ldots + t^{1-G}$. In particular, for 
$G=1$ we have $z(Y) \= 1 \+ Y$ and recover the known 
fermionic Y-system. The bosonic case ($G \= \infty$) 
corresponds to $z(Y) \= Y^{2}/(Y \- 1)$. And the case of 
$G \= 2$ seems to be particularly interesting since here 
$z(Y) \= 1+Y+Y^{-1}$ possesses an additional symmetry 
$z(Y) \= z(Y^{-1})$ or, equivalently, 
$z(\epsilon_a(\theta)) \= z(-\epsilon_a(\theta))$.

\section{Ultraviolet limit}

In the small $r$ limit the scaling function $c(r)$ becomes the
effective central charge of a conformal field theory \cite{Cardy}
which describes the ultraviolet limit of a given massive model.
That is, 
$\lim_{r\rightarrow 0}c(r)=c_{\rm eff} \equiv c-24h^\prime$,
where $c$ is the conformal anomaly and $h^\prime$ is the lowest 
conformal weight in the corresponding conformal field theory.

In order to evaluate $c(r)$ in the $r\rightarrow 0$ limit we
substitute the derivative of (\ref{GTBA}) into (\ref{c(r)}),
replacing in both equations $\cosh \theta$ by $\frac 12 e^\theta$.
Then we integrate the resulting equation by parts, assuming that
$\varphi_{ab}(\theta)$ is symmetric and falls off sufficiently 
fast (typically, $\varphi_{ab}(\theta)= O(e^{-\theta})$ as 
$\theta\rightarrow\infty$). Next we substitute back equation 
(\ref{GTBA}). Finally we change variables from $\theta$ to 
$x(\theta)$, taking into account that $x(\infty)=0$ (as follows 
{}from (\ref{GTBA})), and again integrate 
by parts. All this yields
\begin{equation}\label{c0}
  c_{\rm eff} = \frac{6}{\pi^2} \sum_{a=1}^l c_a \,,
\end{equation}
\begin{equation}\label{c2}
 c_a = - \frac{1}{2} \ln x_a \ln f_a (x_a)
 + \int_0^{x_a} \frac{ {\rm d}t }{t} \ln f_a(t) \,.
\end{equation}
Here we denoted $x_a \equiv x_a(\theta \= 0)$. These quantities
can be determined from the {\em constant} TBA equations that 
follow {}from (\ref{GTBA}) when $r \rightarrow 0$ (this derivation
requires that $x_a(\theta)$ become constant in a region of
$\theta$ of order $-\ln r$ when $r$ is small; see more detailed 
explanations in \cite{TBAKM,CK})
\begin{equation}\label{cTBA}
  \ln x_a = - \sum_{b=1}^l N_{ab} \ln f_b (x_b) \,,
  \qquad  a=1 ,\ldots ,l \,,
\end{equation}
where $2\pi N_{ab}=
 -\int_{-\infty}^{\infty} {\rm d}\theta \varphi_{ab}(\theta)$.

As a simple example we consider the statistics of $\gamma$-ons
(\ref{fga}), assuming for simplicity that $\gamma$ is the same 
for all species (generalization to different $\gamma_a$ is 
obvious). Here the integral in (\ref{c2}) can be computed
explicitly in terms of the Euler dilogarithm which is defined
for $t\leq 1$ as $Li_2(t)= \sum_{k\geq 1} t^k/k^2$. We have
$$
 \int_0^{x} \frac{ {\rm d}t }{t} \ln f_{\gamma}(t) =
 \frac{1}{\gamma} \int_0^{x} \frac{ {\rm d}t }{t}
 \ln ( 1 +\gamma t)  = - \frac{1}{\gamma} \sum_{k=1}^{\infty} 
 \frac{(-\gamma x)^k}{k^2} = -\frac{1}{\gamma} Li_2(-\gamma x) 
$$
and hence 
\begin{equation}\label{cga1}
c_a = -\frac{1}{\gamma} Li_2(-\gamma x_a) 
 - \frac{1}{2\gamma} \ln x_a \ln (1+\gamma x_a) \,,
\end{equation}
where $x_a$ satisfies (\ref{cTBA}). Since $Li_2(-t)=
 \frac 12 Li_2(t^2) - Li_2(t)$,  we can rewrite (\ref{cga1}) 
for positive $\gamma$ employing the Rogers dilogarithm
which is defined as 
$L(t) = Li_2(t) + \frac 12 \ln t \ln (1-t) $ for 
$0 \!\leq\! t \!\leq\! 1$. Then, using the Abel identity 
$L(t^2)=2L(t)-2L(\frac{t}{1+t})$, we obtain
\begin{equation}\label{cga2}
 c_a = \frac{1}{\gamma} 
 L\Bigl( \frac{\gamma x_a}{1+\gamma x_a} \Bigr) 
 + \frac{1}{2\gamma} \ln \gamma \ln (1+\gamma x_a) \,.
\end{equation}
Notice that this expression does not diverge for small $\gamma$;
$\lim_{\gamma\rightarrow 0} c_a= x_a(1- \frac 12 \ln x_a)$ 
is the value corresponding to the Boltzmann statistics.

Consider now the Gentile statistics, assuming that the order $G$ 
is the same for all species. In this case we also can compute the 
integral in (\ref{c2}) explicitly:
$$
 \int_0^{x} \frac{ {\rm d}t }{t} \ln f_{\scriptscriptstyle G}(t) 
 = \int_0^{x} \frac{ {\rm d}t }{t}
 \ln \Bigl( \frac{1-t^{G+1}}{1-t} \Bigr)
 =\sum_{k=1}^{\infty} \frac{1}{k^2}  \Bigl( x^k - 
 \frac{ x^{k(G+1)} }{ G+1}  \Bigr) =  Li_2(x) - 
 \frac{ Li_2 \bigl( x^{(G+1)} \bigr) }{ G+1 } \,,
$$
and we see that for the Gentile statistics expression 
(\ref{c2}) acquires the following nice form involving only 
the Rogers dilogarithms:
\begin{equation}\label{cG}
 c_a = L \left( x_a \right) - {\textstyle \frac{1}{ G+1} }
  L \left( x_a^{G+1} \right) \,.
\end{equation}
Introducing, in agreement with (\ref{Id1}), the constants
$\epsilon_a= -\ln x_a$, we obtain from (\ref{cG}) 
for $G=\infty$ and $G=1$
\begin{equation}\label{cbcf}
 c_a= L(e^{-\epsilon_a}) \qquad {\rm and}  \qquad 
 c_a= L(e^{-\epsilon_a}) - {\textstyle \frac 12 }
 L( e^{-2\epsilon_a}) \,.
\end{equation}
These are the well-known values for the bosonic and 
fermionic statistics \cite{TBAKM}. 
Notice that in the latter case one usually derives
$c_a= L(\frac{1}{1+ e^{\epsilon_a} })$.
This is equivalent to our formula (\ref{cbcf}) due to the Abel 
identity. In fact, the Abel identity can be used in a similar 
way for any $G$ of the form $G = 2^m \- 1$. In this case we
can rewrite (\ref{cG}) as follows
\begin{equation}\label{c2G}
 c_a = \sum_{k=1}^{ m } \frac{1}{ 2^{k-1} } \,  L \Bigl(
 \frac{ (x_a)^{2^{k-1}} }{ 1+ (x_a)^{ 2^{k-1} } } \Bigr) \,.
\end{equation}
In this form $c_a$ is a manifestly positive and monotonic 
function of $x_a$ (the dilogarithm $L(t)$ grows monotonically 
for $0 \!\leq\! t \!\leq\! 1$). Actually, this property holds 
not only for the Gentile statistics but in general case as 
well. To prove this statement we take a derivative of 
(\ref{c2}) and use (\ref{xeq}) and the formula 
(\ref{lim}) for the entropy density. This yields
\begin{equation}\label{dc}
 \partial_{x_a} c_a = \frac{ s(\mu_a(x_a)) }{2 x_a} \,.
\end{equation}
As we discussed in Section 2, we should consider only
such models where $s(\mu(x)) \> 0$. In this case 
the r.h.s.~of (\ref{dc}) is positive.  Hence, 
$c_a$ is always a monotonically increasing function of $x_a$ 
(and therefore a monotonically decreasing function of 
$\epsilon_a$). Moreover,
taking into account (\ref{posit}), we infer from (\ref{c2}) 
that $c_a(0) \= 0$, thus we also proved positivity of 
$c_a(x_a)$. Since these properties are crucial for a physical 
interpretation of $c_a$, let us underline that they hold only 
if the first condition in (\ref{posit}) is fulfilled.
Let us notice also that the properties of $c_a$ together
with (\ref{cTBA}) imply that for any model such that 
$N_{ab} \geq 0$ (which is the case, for instance, for 
all the ADE-affine Toda models \cite{TBAKM})
the value of $c_a$ does not exceed that of
the corresponding free model (where all $N_{ab} = 0$ and 
consequently all $x_a =1$). Hence, (\ref{c2}) evaluated
for $x_a =1$ yields the upper bound on $c_a$,
\begin{equation}\label{cmax}
 c_a \leq  \int_0^1 \frac{ {\rm d}t }{t} \ln f_a(t) \,.
\end{equation}
Notice, however, that for a statistics of the type IIIb
with $R \< 1$ we have $x_a \< 1$, so that in this 
case the upper bound on $c_a$ will be lower than 
(\ref{cmax}). {}From a physical point of view this implies 
that such a statistics cannot emerge in models with weak 
interaction and, in particular, in a free model.

It was argued in \cite{TBAKM} that the quantity 
$\frac{6}{\pi^2} c_a$ (for the ordinary statistics) can 
be interpreted as a massless degree of freedom associated
to the $a$-th species in the massive model. The properties 
of $c_a$ which we proved above show that the interpretation 
of $\frac{6}{\pi^2} c_a$ can be extended to the case of a
generalized extensive statistics as well. 
In this context, it is important to remark that for the 
ordinary statistics (actually, for the Gentile statistics of 
any order) we have $\frac{6}{\pi^2}c_a \leq 1$ as follows 
{}from (\ref{cG}) (recall that $L(1) \= Li_2(1) \= \pi^2/6$) 
if all $N_{ab}$ are non-negative. However, for a general 
statistics the upper bound (\ref{cmax}) on $\frac{6}{\pi^2} c_a$ 
depends on the choice of $f_a$ and may not equal to one. 
For instance, taking $f_a(t)=(1-t)^{-m}$ with $m > 1$, we 
obtain $\frac{6}{\pi^2} c_a = m$ for the corresponding free 
theory. In such a case, the massless degree of freedom of the
corresponding free particle exceeds that of a free boson. 
This can possibly imply that we have to restrict the choice 
of statistics to such $f_a$ that $c_a \leq \frac{\pi^2}{6}$.

\section{Examples related to affine Toda models} 

\subsection*{\normalsize\rm\em \thesection.1 \quad
   Ising and Klein-Gordon models }
The most simple examples which illustrate the features outlined
above more concretely are the Ising model ($A_{1}$-minimal affine 
Toda field theory) with $S(\theta )=-1$ and the Klein-Gordon 
model with $S(\theta )=1.$ In both cases equation (\ref{GTBA}) 
is solved trivially, $ x(\theta )= \exp \{ - rm\cosh \theta \} $.
Then for the Gentile statistics (\ref{Gf}) we can compute the 
entire scaling function (\ref{c(r)}):
\begin{eqnarray}
 c_{\scriptscriptstyle G}(r) &=& 
 \frac{6rm}{\pi^2} \int_0^{\infty} {\rm d}\theta \,
 \cosh \theta \, \Bigl( \ln(1- e^{-(G+1) \, rm\cosh \theta} ) -
 \ln(1- e^{-rm\cosh \theta}) \Bigr)  \nonumber \\
 &=& \frac{6rm}{\pi^2} \sum_{k=1}^{\infty} \frac{1}{k}
 \Bigl( K_{1}(krm) - K_{1}((G+1) krm) \Bigr) \,,  \label{cf}
\end{eqnarray}
where $K_{1}(t)$ is the modified Bessel function. 
We depict $c_{\scriptscriptstyle G}(r)$ in Fig.~2, 
referring to them by a slight abuse of notation as $A_1$. 
One observes that the difference in statistics affects most 
severely the ultraviolet region. The scaling functions for 
different $G$ converge relatively fast towards each other in 
the infrared regime. These appear to be common features of a 
general deformed statistics.

As seen from Fig.~2, the behaviour of
$c_{\scriptscriptstyle G}(r)$ in the ultraviolet region
(there is no first order term in the small $r$ expansion
of $c_{\scriptscriptstyle G}(r)$)
is similar for any finite $G$ to that of the fermionic 
scaling function and is different from the bosonic case
(where the first order term is present). To explain this 
we observe that, as follows from (\ref{cf}),
\begin{equation}\label{cb}
 c_{\scriptscriptstyle G}(r) = c_{\rm b}(r) - 
 {\textstyle \frac{1}{G+1} } c_{\rm b} ((G+1)r) \,, 
\end{equation} 
where $c_{\rm b}(r)$ is the bosonic scaling function. 
Now it is obvious that the first order terms on the
r.h.s.~of (\ref{cb}) cancel each other for any finite $G$.

The well-known property of the asymptotic behaviour of the
modified Bessel function, $\lim_{t\rightarrow 0} tK_{1}(t)=1$,
allows us to derive from (\ref{cf})
\begin{equation}\label{cGt}
 c^{\rm eff}_{\scriptscriptstyle G} = 
 \lim_{r\rightarrow 0} c_{\scriptscriptstyle G}(r)=
 \frac{6}{\pi^2} \sum_{k=1}^{\infty} \frac{1}{k^2}
 \Bigl( 1 - \frac{1}{G+1} \Bigr) = \frac{G}{G+1} \,,
\end{equation}
which is in agreement with (\ref{cG}) since $x_a \= 1$ now.
It is worth noticing that $c^{\rm eff}_{\scriptscriptstyle G}$ 
can be identified for every positive integer $G$ as the effective 
central charge of a minimal conformal model ${\cal M}(s,t)$ 
(such that $st=6(G+1)$). In particular, $G=1,2,3,4$ correspond 
to the ${\cal M}(3,4)$, ${\cal M}(2,9)$, ${\cal M}(3,8)$, 
${\cal M}(5,6)$ minimal models, respectively.

For the $\gamma$-ons (\ref{fga}) the scaling function related 
to the $A_1$-models can also be computed:
\begin{equation}\label{cfga}
 c_{\gamma}(r) =  \frac{6rm}{\gamma\pi^2} \int_0^{\infty} 
 {\rm d}\theta \,  \cosh \theta \, 
 \ln (1+ \gamma e^{- rm\cosh \theta} ) =
 \frac{6 rm}{ \pi^2} \sum_{k=1}^{\infty} \frac{1}{k}
  (-\gamma)^{k-1} K_{1}(krm) \,.
\end{equation}
Using again the asymptotic behaviour of the modified Bessel 
function, we obtain 
$c^{\rm eff}_{\gamma} = \lim_{r\rightarrow 0} c_{\gamma}(r)=
 - \frac{6}{ \gamma \pi^2} Li_2(-\gamma)$ in agreement with
(\ref{cga1}). Here, unlike in the Gentile case, it is not 
easy to see whether the effective central charge 
$c^{\rm eff}_{\gamma}$ has rational value for a given $\gamma$. 
As we discussed in Section 2, positive $\gamma$ is to be of 
the form $\gamma = 1/d$ with $d$ positive integer. It appears 
then that the only $d$ leading to rational value of 
$c^{\rm eff}_{\gamma}$ is $d=1$. On the other hand, $\gamma$
varies continuously in the range $-1 \leq \gamma \leq 0$, so 
that any rational value of $c^{\rm eff}_{\gamma}$ not 
exceeding $\frac{6}{\pi^2}$ occurs here. For the
Boltzmann statistics $c^{\rm eff}_{0}=\frac{6}{\pi^2}$
is irrational. Finally, $\gamma \< -1$ is the case of 
a type IIIb statistics with $R \< 1$, which, as we 
discussed above, cannot emerge in a free model.

\subsection*{\normalsize\rm\em \thesection.2 \quad
    Scaling Potts and Yang-Lee models}

Next we consider the scaling Potts model, which was studied 
previously in the TBA framework for the fermionic statistics in 
\cite{Zam} and for the Haldane-Wu statistics in \cite{BF}. 
The two particles in the model are conjugate to each other and 
consequently their masses are the same, $m_{1} \= m_{2} \= m$. 
The conjugate particle occurs as a bound state when two particles 
of the same species scatter. The S-matrix of the scaling Potts 
model equals the minimal S-matrix of the $A_{2}$-affine Toda 
field theory. The corresponding quantities $\varphi_{ab}(\theta)$ 
that enter the TBA equation are given by
\begin{equation}\label{SPphi}
 \varphi_{11}(\theta) = \varphi_{22}(\theta) =
 \frac{-\sqrt{3}}{2\cosh \theta +1} \qquad  {\rm and} \qquad 
 \varphi_{12}(\theta) = \frac{\sqrt{3}}{1-2\cosh \theta } \,.
\end{equation}

Consider the scaling Potts model with both species of particles
obeying the Gentile statistics of the same order $G$. In this 
case the $Z_{2}$-symmetry of the model is preserved so that 
$x_{1}(\theta )=x_{2}(\theta) \equiv x(\theta)$ and (\ref{GTBA}) 
reduces to a single integral equation. It appears to be impossible 
to find an analytic solution $x(\theta)$ to this equation, but it 
is straightforward to solve it numerically. Taking 
$x^{[0]}(\theta ) = \exp\{-rm\cosh \theta\} $ as the first 
approximation, we can iterate (\ref{GTBA}) as follows
\begin{equation}\label{a2}
 \ln \left( x^{[n+1]}(\theta )\right) = -rm \cosh \theta -
 \frac{2\sqrt{3}}{\pi} \int_{-\infty }^{\infty}
 {\rm d}\theta^\prime \frac{\cosh (\theta -\theta^\prime)}{
 1+2 \cosh 2(\theta -\theta^\prime)}  
 \ln \left( f_{\scriptscriptstyle G} 
 \bigl( x^{[n]}(\theta^\prime) \bigr) \right).
\end{equation}
Convergence is achieved relatively quickly (it depends on the 
value of $mr$). The results for $G \= 1$ (which is in complete
agreement with the calculation in \cite{Zam}) and for $G \= 2$ 
are shown in Fig.~3. To make contact with the literature, we 
introduced the quantity 
$L(\theta )=\ln f_{\scriptscriptstyle G}(x(\theta))$.
One observes the typical plateau $L(\theta)=const$ for some region 
of $\theta$ when $mr \rightarrow 0$, which is required to derive 
(\ref{cTBA}). We see also that for the same value of $mr$ the 
functions $L(\theta)$ corresponding to different $G$ have similar 
profiles but different heights of the plateau.

Having solved the generalized TBA equation, we can substitute
$x(\theta)$ into (\ref{c(r)}) and compute the entire scaling 
function $c_{\scriptscriptstyle G}(mr)$. The result of the 
numerical computation is shown in Fig.~4. The behaviour of
the scaling functions is similar to what we have already 
observed in the $A_1$ case.

Due to the fact that the scattering matrices of the scaling Potts 
model and the scaling Yang-Lee model are related as 
$S^{YL}(\theta) = S_{11}^{A_{2}}(\theta )S_{12}^{A_{2}}(\theta )$, 
the generalized TBA equations for the two models coincide. 
The only difference is that the scaling Yang-Lee model has only one 
particle. Therefore its scaling function equals a half of that 
for the scaling Potts model.

\subsection*{\normalsize\rm\em \thesection.3 \quad
    Comparison with Haldane-Wu statistics }

In Section 3 we have seen that the entropy density corresponding
to the Gentile statistics majorizes that for the Haldane-Wu
statistics. It appears (at least for certain models) that an
analogous relation holds for the corresponding finite-size scaling 
functions. Specifically, if $c_{\scriptscriptstyle G}(r)$ is the 
scaling function (\ref{c(r)}) of $A_1$ or $A_2$ minimal affine
Toda model where all species obey order $G$ Gentile statistics 
and $c_g(r)$ is the scaling function (\ref{hwcr}) of the same 
model where the Haldane-Wu statistical interaction is of the 
form $g_{ab} = \frac{1}{G} \delta_{ab}$, 
then for all $r \geq 0$ we have
\begin{equation}\label{crin}
 c_{\scriptscriptstyle G}(r) > c_g(r) \,, \qquad {\rm where}
 \qquad g_{ab} = {\textstyle \frac{1}{G} } \delta_{ab} 
 \quad {\rm and} \quad G>1 \,.
\end{equation}
For the $A_1$ case we can rigorously prove relation 
(\ref{crin}) since the solutions of the generalized TBA
equations (\ref{GTBA}) and (\ref{hwTBA}) are found explicitly:
\begin{equation}\label{xy0}
 g \ln y(\theta) + (1-g) \ln (1+y(\theta)) = 
 mr \, \cosh\theta = - \ln x(\theta) \,.
\end{equation}
This equation allows us to express $x(\theta)$ in terms of 
$y(\theta)$:
\begin{equation}\label{xy1}
 x(\theta)  = y^{-g}(\theta) \, (1+y(\theta))^{g-1} \,.
\end{equation}
Moreover, since $mr \cosh\theta \> 0$, we infer 
{}from (\ref{xy0}) that
\begin{equation}\label{yy0}
 y(\theta) > y_0 \,, \qquad {\rm where} \qquad
  y_0^g = (1+y_0)^{g-1} \,.
\end{equation}
As seen from (\ref{c(r)}) and (\ref{hwcr}), in order to 
establish (\ref{crin}) it suffices to show that 
$f_{\scriptscriptstyle G =1/g}(x(\theta)) > 1+y^{-1}(\theta)$.
The latter relation is equivalent to the inequality
\begin{equation}\label{yy}
 1- (1+y(\theta))^{-\frac 1g} > 
 y^g(\theta) \, (1+y(\theta))^{-g} \,,
\end{equation}
as can be verified by simple algebraic manipulations using 
formula (\ref{xy1}). Now it is elementary to check that 
(\ref{yy}) is valid provided that (\ref{yy0}) holds. Thus, 
we proved the assertion (\ref{crin}) in the $A_1$ case. For 
illustration, the scaling function $c_g(r)$ for $g \= 1/2$ 
Haldane-Wu statistics is shown in Fig.\ 2. It lies everywhere 
below $c_{\scriptscriptstyle G}(r)$ for $G \= 2$ Gentile 
statistics with the maximal difference reached 
in the ultraviolet region.
 
For a higher rank case, already for $A_2$, the generalized TBA
equations are integral and it is not clear whether we can compare
$x_a(\theta)$ and $y_a(\theta)$ analytically. However, numerical
computations for various $G$ show that (\ref{crin}) holds in 
the $A_2$ case as well. For illustration, we present in Fig.\ 4
the scaling function $c_g(r)$ for the Haldane-Wu statistics 
with $g_{ab} \= \frac 12 \delta_{ab}$.

A particular consequence of (\ref{crin}) is the following
inequality for the corresponding central charges:
\begin{equation}\label{cin}
 c^{\rm eff}_{\scriptscriptstyle G} > c^{\rm eff}_g 
 \,, \qquad {\rm for}  
 \quad g_{ab} = {\textstyle \frac{1}{G} } \delta_{ab} 
 \quad {\rm and} \quad G>1 \,.
\end{equation}
Here $c^{\rm eff}_{\scriptscriptstyle G}$ is found {}from 
(\ref{c0}), (\ref{cTBA}) and (\ref{cG}) whereas 
$c^{\rm eff}_g$ is the effective central charge related 
to the Haldane-Wu statistics with 
$g_{ab} = \frac{1}{G} \delta_{ab}$ and is given by \cite{BF}
\begin{equation}\label{chw1}
  c^{\rm eff}_g = \frac{6}{\pi^2}  \sum_{a=1}^l 
  L\Bigl( \frac{1}{1+y_a} \Bigr) \,, \qquad {\rm where}
 \quad  \ln(1+y_a) = \sum_{b=1}^l ( N_{ab} + g \delta_{ab} )
  \ln (1+y_a^{-1}) \,. 
\end{equation}
Again, in the $A_1$ case we can prove relation (\ref{cin}) 
directly. Here $c^{\rm eff}_{\scriptscriptstyle G}$ has a 
simple form (\ref{cGt}) and therefore (\ref{cin}) reduces to
\begin{equation}\label{din1}
 \frac{1}{1+g} > \frac{6}{\pi^2} 
 L\Bigl( \frac{1}{1+y_0} \Bigr) \,,
\end{equation}
where $y_0$ is defined as in (\ref{yy0}). Notice that 
$g \< y_0 \< 1$ for $g\neq 0,1$ (since $y_0 \leq g$ 
would contradict (\ref{Gin})).
Therefore, $\frac 12 \< \frac{1}{1+y_0} \< \frac{1}{1+g}$.
Using now a property of the dilogarithm, 
\begin{equation}\label{Lin}
 L(t) < \frac{\pi^2}{6} t  \qquad {\rm for} \quad 
 {\textstyle \frac 12 < t < 1 } \,; \qquad
 L(t) > \frac{\pi^2}{6} t  \qquad {\rm for} \quad 
 {\textstyle 0 < t <  \frac 12 } \,, 
\end{equation}
we establish the desired inequality (\ref{din1}).

As we discussed above, the order $G$ of the Gentile 
statistics is to be a positive integer. However, if we 
regard (\ref{c0}), (\ref{cTBA}) and (\ref{cG}) as 
a {\em definition} of the quantity
$c^{\rm eff}_{\scriptscriptstyle G}$, then (\ref{din1})
extends to any $G \>1 $ (since the proof does not require 
that $G$ be an integer) as shown in Fig.~5. 
Furthermore, considering $G$ in 
the range $0 \< G \< 1$, we can prove with the help of 
(\ref{Gin}) that here $y_0 < g$ and hence
$\frac{6}{\pi^2} L(\frac{1}{1+y_0}) \> \frac{1}{1+g}$
due to (\ref{Lin}). Thus, for this range of $G$ we 
have a {\em reverse} inequality, i.e., 
\begin{equation}\label{cin2}
 c^{\rm eff}_{\scriptscriptstyle G} < c^{\rm eff}_g
 \,, \qquad {\rm for}  
 \qquad g_{ab} = {\textstyle \frac{1}{G} } \delta_{ab} 
 \quad {\rm and} \quad 0<G<1 \,.
\end{equation}
As we will see below, this inequality holds for higher
rank cases as well. But let us stress again that 
$c^{\rm eff}_{\scriptscriptstyle G}$ in (\ref{cin2})
is a formally defined quantity which is not related 
directly to the Gentile statistics.

The inequality for the central charges (\ref{cin}) is a
weaker statement than the inequality for the scaling 
functions (\ref{crin}). Furthermore, (\ref{cin}) involves 
dilogarithms, which makes its proof more complicated 
(already in the $A_1$ case we used a rather subtle 
property (\ref{Lin})). But the advantage of (\ref{cin}) 
for a numerical verification is that it relates numbers 
and not functions as (\ref{crin}). Moreover, since 
$c(r)$ is a continuous function, validity of (\ref{cin}) 
implies that (\ref{crin}) holds at least in some 
ultraviolet region. Therefore, we will discuss below
validity of (\ref{cin}) for several affine Toda models.
As a mathematical by-product, this provides us with
interesting dilogarithm inequalities.

In the $A_2$ case, eq.~(\ref{cin}) contains dilogarithms in a 
more involved way. As seen {}from (\ref{cTBA}), (\ref{cG}) and
(\ref{chw1}), it is equivalent to the following inequality
\begin{equation}\label{din2}
 L ( \tilde{x}_0 ) - \frac{g}{1+g} 
 L\Bigl( \tilde{x}_0^{1+\frac 1g} \Bigr) > 
 L\Bigl(\frac{1}{ 1+\tilde{y}_0 } \Bigr) \,,
  \qquad {\rm for} \quad 0 < g < 1 \,.
\end{equation}
Here $\tilde{x}_0$ and $\tilde{y}_0$ are determined {}from 
the following equations that follow from (\ref{cTBA}) and 
(\ref{chw1}) upon noticing that $N_{11} \+ N_{12} \= 1$
\begin{equation}\label{din3}
 \tilde{x}_0 \, f_{1/g} ( \tilde{x}_0) = 1  \qquad 
 {\rm and} \qquad  \tilde{y}_0^{1+g} = ( 1+\tilde{y}_0 )^g \,.
\end{equation}
It is worth remarking that, as was noticed in \cite{BF},
$\frac{6}{\pi^2} L(\frac{1}{ 1+ \tilde{y}_0} )$ coincides with 
the effective central charge of the Calogero-Sutherland 
model with the coupling constant $\lambda =g$. 

Numerical computations show that (\ref{din2}) indeed holds
for any $0 \< g \< 1$, that is for any $G \>1$ (here again 
we need not to restrict $G$ to be an integer). Furthermore, 
like in the $A_1$ case, inequality (\ref{din2}) reverses for 
$0 \< G \< 1$, i.e., for $g \> 1$. For illustration, we 
depict the corresponding $c^{\rm eff}_{\scriptscriptstyle G}$ 
and $c^{\rm eff}_g$ in Fig.~5. This result provides an 
additional support to our claim that in the $A_2$ case the 
entire scaling function $c_{\scriptscriptstyle G}(r)$ 
majorizes $c_g(r)$ for $G \> 1$.

It is interesting to remark that, as follows from (\ref{cG})
and (\ref{chw1}), the bosonic version ($G \= \infty$ or, 
equivalently, $g \= 0$) of the $A_2$-minimal affine Toda 
model has $c_a \= \frac 12$. That is, then each species 
in the scaling Potts model (or the only species in the 
scaling Yang-Lee model) appears to be a free fermion, 
whereas the entire central charge is that of a free boson
(see also related comments in \cite{BF}). That is why
all curves on Fig.~5 converge to the same value $c \= 1$.
 
\subsection*{\normalsize\rm\em \thesection.4 \quad
    Higher rank cases }
 
Now we discuss briefly the simplest minimal affine Toda 
models which have particle species with different masses. 
We will compare numerical values of 
$c^{\rm eff}_{\scriptscriptstyle G}$ and $c^{\rm eff}_{g}$ 
again setting $g_{ab} \= \frac 1G \delta_{ab}$. 
For this purpose we use (\ref{cTBA}) and (\ref{chw1}),
taking into account \cite{TBAKM} that the $N$-matrix in 
these equations is given by 
$N_{\rm\bf g} \= I_{\rm\bf g} (2-I_{\rm\bf g})^{-1}$, 
where $I_{\rm\bf g}$ stands for the incidence matrix of 
the corresponding Lie algebra {\bf g}.

In the $A_3$ case the masses of the three species are 
$m_1 \= m_3 \= m_2/\sqrt{2}$.
Then, computing $c^{\rm eff}_{\scriptscriptstyle G}$
by formulae (\ref{cTBA}) and (\ref{cG}), we find:
$c^{\rm eff}\approx 0.77$, $c^{\rm eff}=1$,
$c^{\rm eff}\approx 1.12$, $c^{\rm eff}\approx 1.16$
for $G=\frac 12, 1, 2, \infty$, respectively.
The values of $c^{\rm eff}_{g}$ corresponding to the
Haldane-Wu statistics with $g_{ab} = \frac{1}{G} \delta_{ab}$
were found in \cite{BF}:
$c^{\rm eff}\approx 0.89$, $c^{\rm eff}=1$,
$c^{\rm eff}\approx 1.07$, $c^{\rm eff}\approx 1.16$
for $g=2, 1, \frac 12, 0$, respectively. 
Carring out analogous computations for the $A_4$ case,
where the masses of the four species are 
$m_1 \= m_4$, $m_2 \= m_3 \= m_1 (\sqrt{5} \+ 1)/2$,
we find (Gentile statistics): 
$c^{\rm eff}\approx 0.92$, $c^{\rm eff}=8/7$,
$c^{\rm eff}\approx 1.25$, $c^{\rm eff}\approx 1.28$
for $G=\frac 12, 1, 2, \infty$, respectively; and
(Haldane-Wu statistics): 
$c^{\rm eff}\approx 1.05$, $c^{\rm eff}=8/7$,
$c^{\rm eff}\approx 1.20$, $c^{\rm eff}\approx 1.28$
for $g=2, 1, \frac 12, 0$, respectively.

One can also consider a folded algebra case with two 
different masses, namely $A_4^{(2)}$, where 
$m_2 \= m_1 (\sqrt{5} \+ 1)/2$. In general, one
assigns to the $A_{2n}^{(2)}$ minimal affine Toda 
model a tad-pole graph \cite{TBAKM,Rav} such that the 
corresponding incidence matrix differs from that of 
$A_n$ only by 1 in the lower-right entry. Then one 
can check that $x_a$ in (\ref{cTBA}) and $y_a$ in 
(\ref{chw1}) coincide with those for the $A_{2n}$ case 
(a well-known fact for the fermionic statistics 
\cite{TBAKM,Zam}). Consequently, for any statistics
we have $c^{\rm eff}(A_{2n}^{(2)}) \= \frac 12
 c^{\rm eff}(A_{2n})$. Therefore, the data for
$A_4^{(2)}$ follow from the above results for $A_4$.

Finally, in the $D_4$ case, where the masses of the 
four species are 
$m_1 \= m_3 \= m_4 \= m_2/\sqrt{3}$, we obtain for
the Gentile statistics: 
$c^{\rm eff}\approx 0.82$, $c^{\rm eff}=1$,
$c^{\rm eff}\approx 1.08$, $c^{\rm eff}\approx 1.09$
for $G=\frac 12, 1, 2, \infty$, respectively; and
for the Haldane-Wu statistics: 
$c^{\rm eff}\approx 0.93$, $c^{\rm eff}=1$,
$c^{\rm eff}\approx 1.04$, $c^{\rm eff}\approx 1.09$
for $g=2, 1, \frac 12, 0$, respectively.

We see that all the above results are in agreement with 
(\ref{cin}) and (\ref{cin2}). This, together with the
more detailed results we obtained in the $A_1$ and $A_2$ 
cases, allows us to conjecture that these inequalities 
and the inequality (\ref{crin}) for the scaling functions
hold actually for all simply-laced minimal affine Toda models.

We remark also that in all the computations of 
$c^{\rm eff}$ in this subsection we always have the
rule that the value of $c_a$ is smaller for the species 
which has heavier particle (this was known for the 
fermionic statistics \cite{TBAKM}). This observation 
provides an additional support to the above discussed 
interpretation of $\frac{6}{\pi^2}c_a$ as a massless 
degree of freedom of the corresponding particle.

\section{Conclusion}

Summarizing, we considered possible types of a generalized 
extensive statistics and established properties of the 
corresponding entropy densities. In particular, we
established numerically (and gave some supporting analytical
arguments) that the entropy density for the Gentile statistics
of order $G \> 1$ majorizes that for the Haldane-Wu 
statistics if they correspond to the same maximal occupancy.
Further, we derived the thermodynamic Bethe ansatz equation 
and the finite-size scaling function $c(r)$ for a relativistic 
multi-particle system obeying a generalized extensive 
exclusion statistics. We put particular emphasis on the 
ultraviolet limit of such a system. We derived an expression 
for an effective central charge in the general case and 
showed that for the Gentile statistics it acquires an elegant 
form involving dilogarithms. We discussed a physical 
interpretation of the `partial' central charges $c_a$
and argued that it possibly leads to restrictions on the 
choice of statistics. Finally, we observed (and partially
proved) that the majorizing properties of the Gentile 
statistics with respect to the Haldane-Wu statistics
extend also to finite-size scaling functions and central 
charges related to the minimal simply-laced affine Toda models. 

In the presented analysis of thermodynamic properties of 
a relativistic multi-particle system the single-state 
partition function $f(t)$ played a key role. For a given 
model, if we know $f(t)$ (at least in the thermodynamic 
limit) and the quantities $\varphi_{ab}(\theta)$, then we 
can compute, at least in principle, the entire scaling 
function which provides complete thermodynamic description
of a system. In principle, $f(t)$ should be determinable
{}from the corresponding Hamiltonian. However, this might
be a non-trivial task since the exclusion statistics
of particles can differ from the exchange statistics of 
the fields present in the Hamiltonian.
Therefore, it may be more practical for a given system to 
fix an exotic statistics a-priory and investigate whether 
it leads to realistic physical properties. In particular,
one can obtain information about the ultraviolet limit
and see if it corresponds to an appropriate conformal 
field theory. As we have seen above, the different types
(\ref{types}) of $f(t)$ lead to significantly different
thermodynamic properties. The question, which of these 
statistics can emerge in physical models, remains open.

Another open problem is to prove rigorously the assertion 
(\ref{ineq2}) that the entropy density for the Gentile 
statistics majorizes that for the Haldane-Wu statistics. 
Such a proof will probably provide a deeper insight into 
general properties of exotic exclusion statistics.
Also it would be interesting to verify the
majorizing inequalities for the scaling functions 
(\ref{crin}) and central charges (\ref{cin}) related
to the affine Toda models and to understand whether
such a majorization is a model independent feature
of the two involved statistics.

\vspace*{1mm}
{\bf Acknowledgements:}
I am grateful to A.\ Fring for proposing the problem and
participation in the initial stage of the research.
This work was supported in part by INTAS grant 99-01459
and by Russian Fund for Fundamental Investigations 
grant 99-01-00101.

\newpage\normalsize

\vspace*{-5mm}
\includegraphics[width=62mm,bburx=80mm,bbury=80mm]{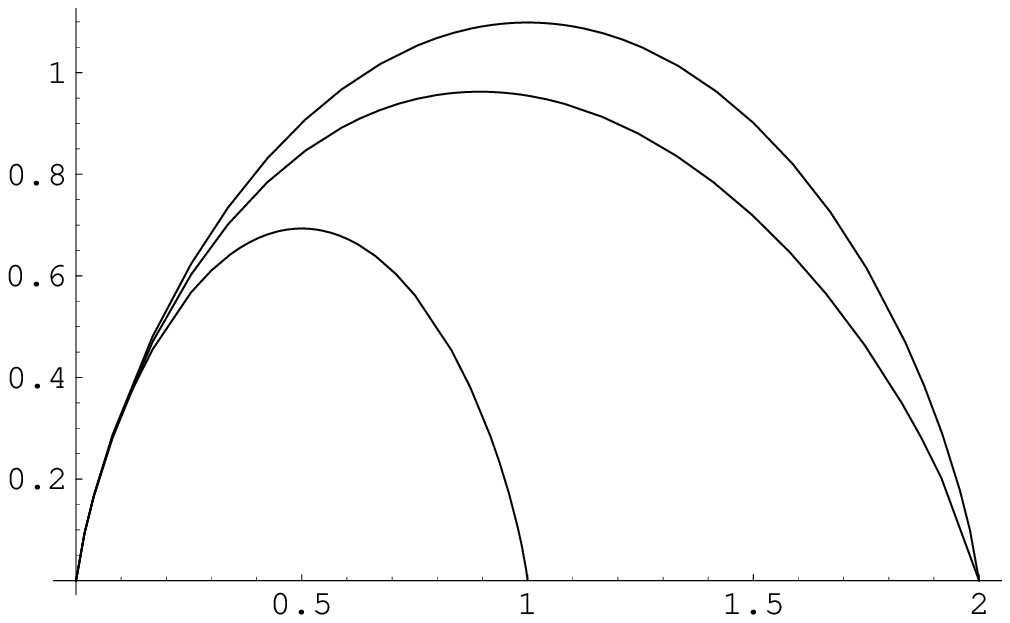}

\vspace*{-32mm} {\large \hspace*{10mm} $s(\mu)$}

\vspace*{16mm} \hspace*{112mm} {\large $\mu$}

\vspace*{5mm}
\noindent {\small
Fig.\ 1: Entropy densities for $G=2$ Gentile statistics (the
upper curve), for $g=1/2$ Haldane-Wu statistics (the middle
curve) and for the fermionic statistics (the lower curve). }

\includegraphics[width=62mm,bburx=80mm,bbury=80mm]{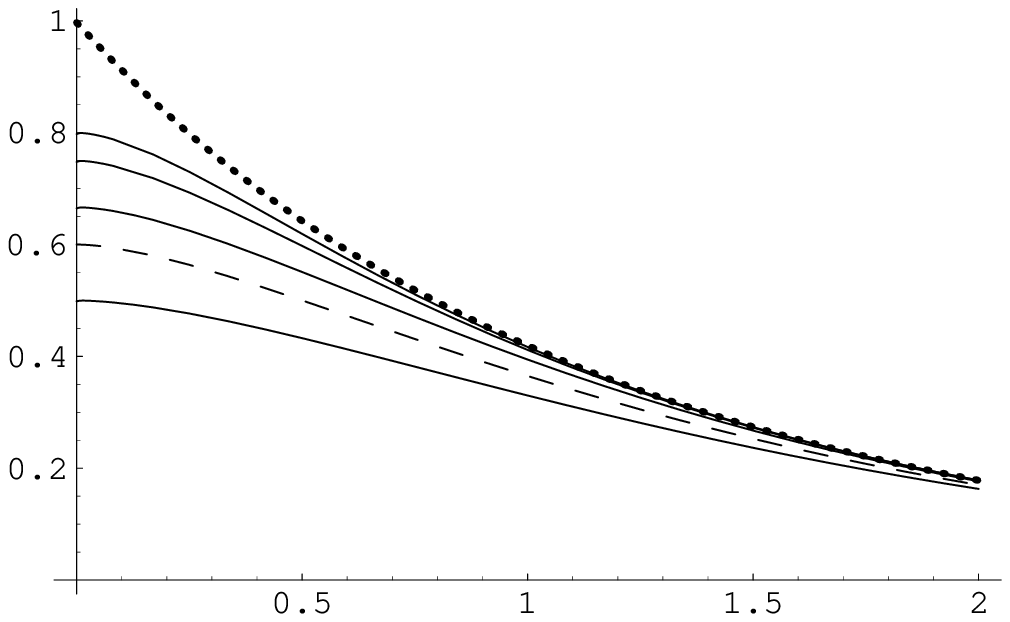}

\vspace*{-32mm} {\large \hspace*{10mm} $c(mr)$}

\vspace*{16mm} \hspace*{112mm} {\large $mr$}

\vspace*{6mm}
\noindent {\small
Fig.\ 2: Scaling functions of the $A_1$-minimal affine Toda model 
for the Gentile statistics of order $G=1,2,3,4$ (solid curves,
{}from down to up), for $g=1/2$ Haldane-Wu statistics (the dashed 
curve) and for the bosonic statistics (the dotted curve). }

\includegraphics[width=62mm,bburx=80mm,bbury=80mm]{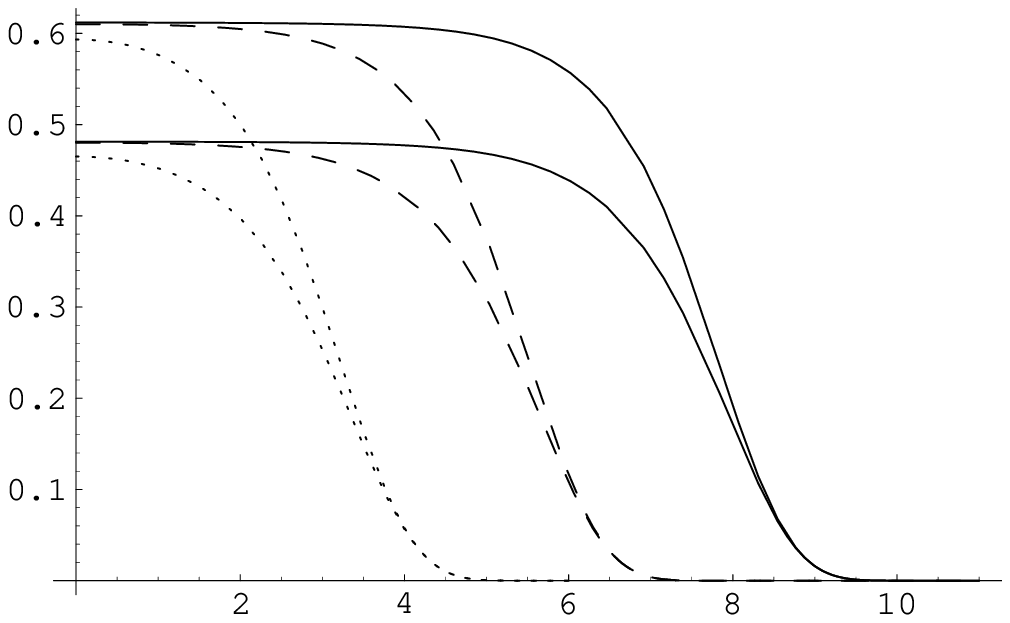}

\vspace*{-32mm} {\large \hspace*{10mm} $L(\theta)$}

\vspace*{16mm} \hspace*{112mm} {\large $\theta$}

\vspace*{6mm}
\noindent {\small
Fig.\ 3: Solution $L(\theta)$ of the TBA equation for the
$A_2$-minimal affine Toda model for the scaling parameter 
$mr=0.1$ (dotted curves), $mr=0.01$ (dashed curves), and 
$mr=0.001$ (solid curves). For the same value of $mr$ the 
lower curve corresponds to the fermionic statistics and the 
upper curve to $G \= 2$ Gentile statistics.}

\newpage

\includegraphics[width=62mm,bburx=80mm,bbury=80mm]{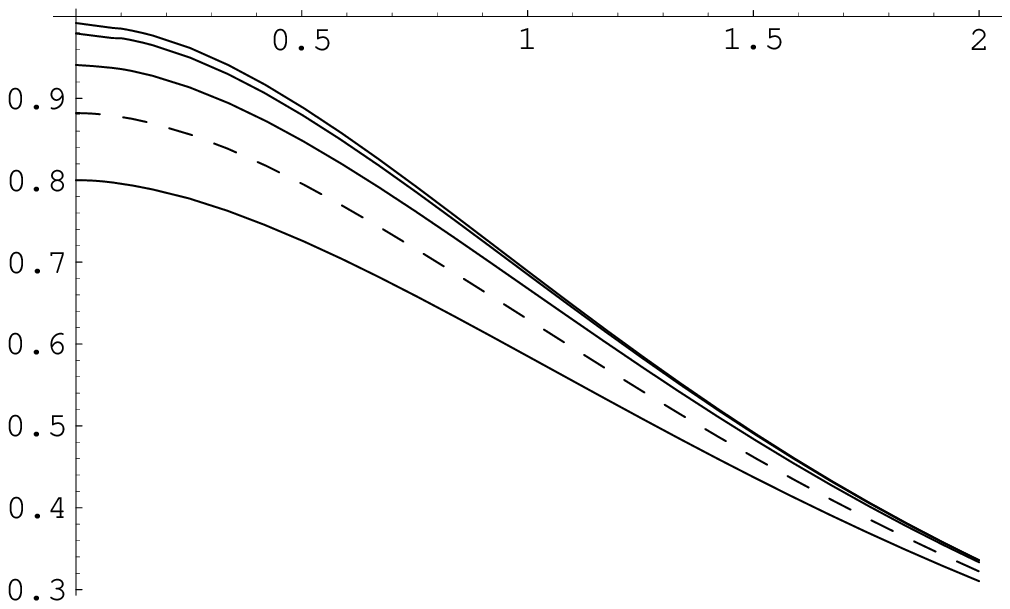}

\vspace*{-30mm} {\large \hspace*{10mm} $c(mr)$}

\vspace*{-22mm} \hspace*{112mm} {\large $mr$}

\vspace*{47mm}
\noindent {\small
Fig.\ 4: Scaling functions of the $A_2$-minimal affine Toda 
model for the Gentile statistics of order $G\ = 1,2,3,4$ (solid 
curves, {}from down to up) and for the Haldane-Wu statistics
with $g_{ab} = \frac 12 \delta_{ab}$ (the dashed curve).}

\includegraphics[width=62mm,bburx=80mm,bbury=80mm]{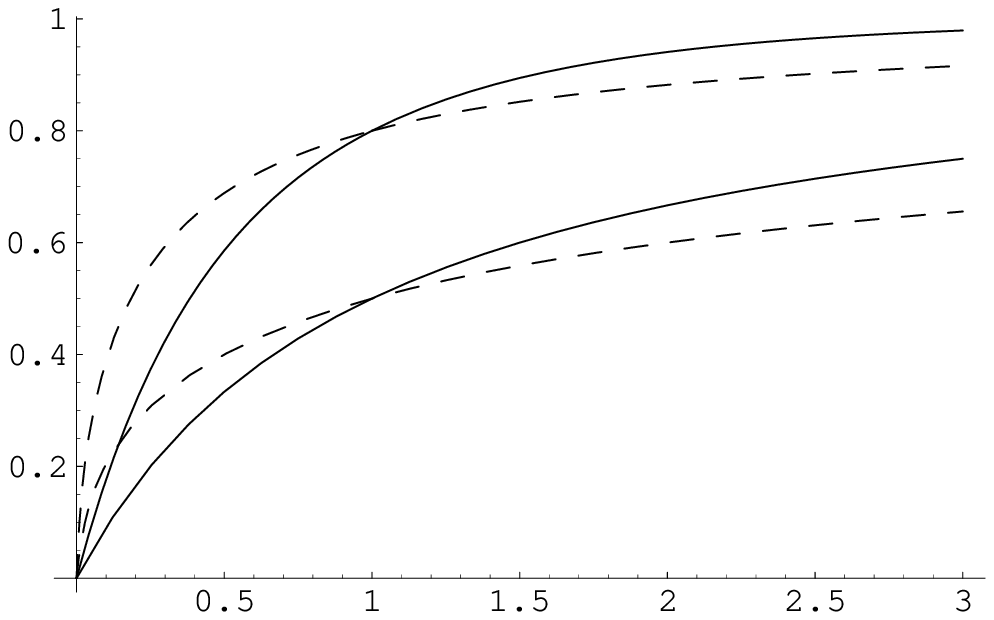}

\vspace*{-32mm} {\large \hspace*{10mm} $c_{\rm eff}$}

\vspace*{16mm} \hspace*{112mm} {\large $G$}

\vspace*{8mm}
\noindent {\small
Fig.\ 5: Effective central charge of the $A_1$ and 
$A_2$-minimal affine Toda models for the Gentile statistics 
of order $G$ (lower and upper solid curves) and for the 
Haldane-Wu statistics with $g_{ab} = \frac{1}{G}  \delta_{ab}$ 
(lower and upper dashed curves).}

\end{document}